\documentclass[fleqn,12pt,twoside]{article}
\usepackage{espcrc1}

\usepackage{graphicx}
\usepackage[figuresright]{rotating}

\newcommand{\AmS}{{\protect\the\textfont2
  A\kern-.1667em\lower.5ex\hbox{M}\kern-.125emS}}

\def\nabstar#1{\nabla\kern-0.5pt\smash{\raise 4.5pt\hbox{$\ast$}}
               \kern-4.5pt_{#1}}

\def\drvstar#1{\partial\kern-0.5pt\smash{\raise 4.5pt\hbox{$\ast$}}
               \kern-5.0pt_{#1}}

\def\newline{\relax\ifhmode\null\hfil\break\else\nonhmodeerr@\newline\fi}
\def\frac#1#2{{#1\over#2}}
\def\text#1{{\hbox{\rm #1}}}

\newcommand{\beq}{\begin{equation}}
\newcommand{\eeq}{\end{equation}}
\newcommand{\bea}{\begin{eqnarray}}
\newcommand{\eea}{\end{eqnarray}}

\def\EQ{\hspace{-2mm} &=& \hspace{-2mm}}

\def\BA{\begin{eqnarray}}
\def\EA{\end{eqnarray}}
\def\BAN{\begin{eqnarray*}}
\def\EAN{\end{eqnarray*}}

\def\gm5{\gamma_5}

\def\qu{{\bf u}}
\def\qd{{\bf d}}
\def\qs{{\bf s}}

\title{Baryon Masses in Lattice QCD with Exact Chiral Symmetry\thanks{
This work was supported in part by the National Science Council (ROC) 
under Grant No. NSC93-2112-M002-016,  
and by the National Science Foundation under Grant No. PHY99-07949.}}  

\author{Ting-Wai Chiu\address[NTU]{Physics Department, 
                                   National Taiwan University,
                                    Taipei, Taiwan.}\address[KITP]{Kavli Institute for Theoretical Physics, University of California, Santa Barbara.} 
        and Tung-Han Hsieh\addressmark[NTU] }

\begin{document}

\maketitle

\begin{abstract}
We investigate the baryon mass spectrum 
in quenched lattice QCD with exact chiral symmetry.
For 100 gauge configurations generated with Wilson gauge action 
at $ \beta = 6.1 $ on the $ 20^3 \times 40 $ lattice, 
we compute (point-to-point) quark propagators for 30 quark masses 
in the range $ 67 \mbox{ MeV} \le m_q  \le 1790 \mbox{ MeV} $. 
For baryons only composed of strange and charm quarks,
their masses are extracted directly from the time correlation functions, 
while for those containing $ u (d) $ light quarks, their masses
are obtained by chiral extrapolation to $ m_\pi = 135 $ MeV.
Our results of baryon masses are in good agreement with experimental
values, except for the negative parity states of  
$ \Lambda $ and $ \Lambda_c $. Further, our results of 
charmed (including doubly-charmed 
and triply-charmed) baryons can serve as predictions of QCD.
\end{abstract}

\section{Introduction}

\begin{figure}[htb]
\begin{minipage}[t]{80mm}
\includegraphics*[height=7cm,width=8cm]{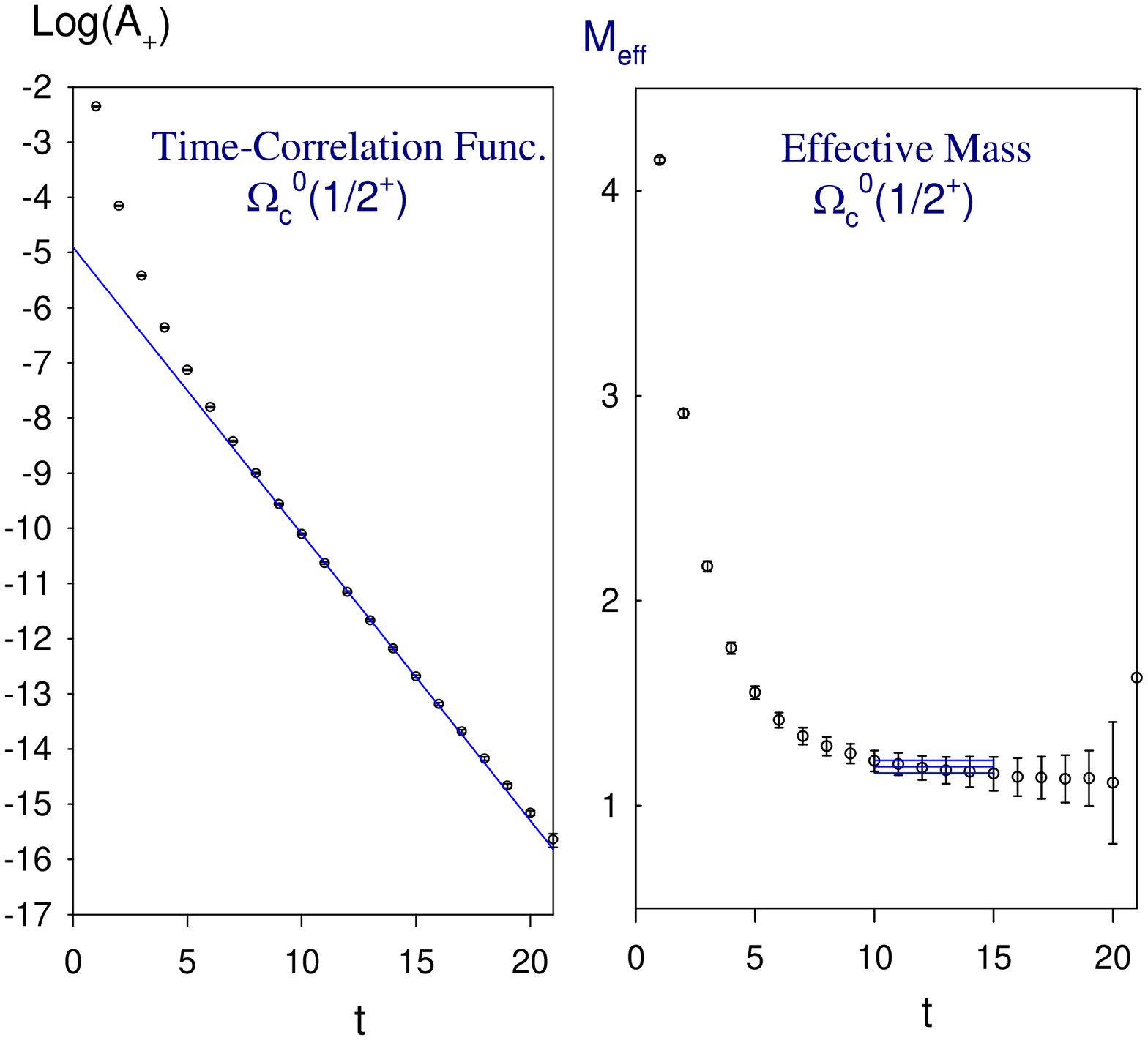}
\caption{The even parity amplitude $ A_+(t) $ of the time correlation
function of $ \Omega_c^0 $ is plotted versus the time slice. The solid line
is the single exponential fit for $ 10 \le t \le 15 $ where the effective
mass attains a plateau, and it gives $ M_{\Omega_{c}^0(1/2^+)}= 2677(30) $ MeV, 
in good agreement with the mass of $ \Omega_c^0 (2697) $.}
\label{fig:omega_c_20}
\end{minipage}
\hspace{\fill}
\begin{minipage}[t]{75mm}
\includegraphics*[height=7cm,width=6cm]{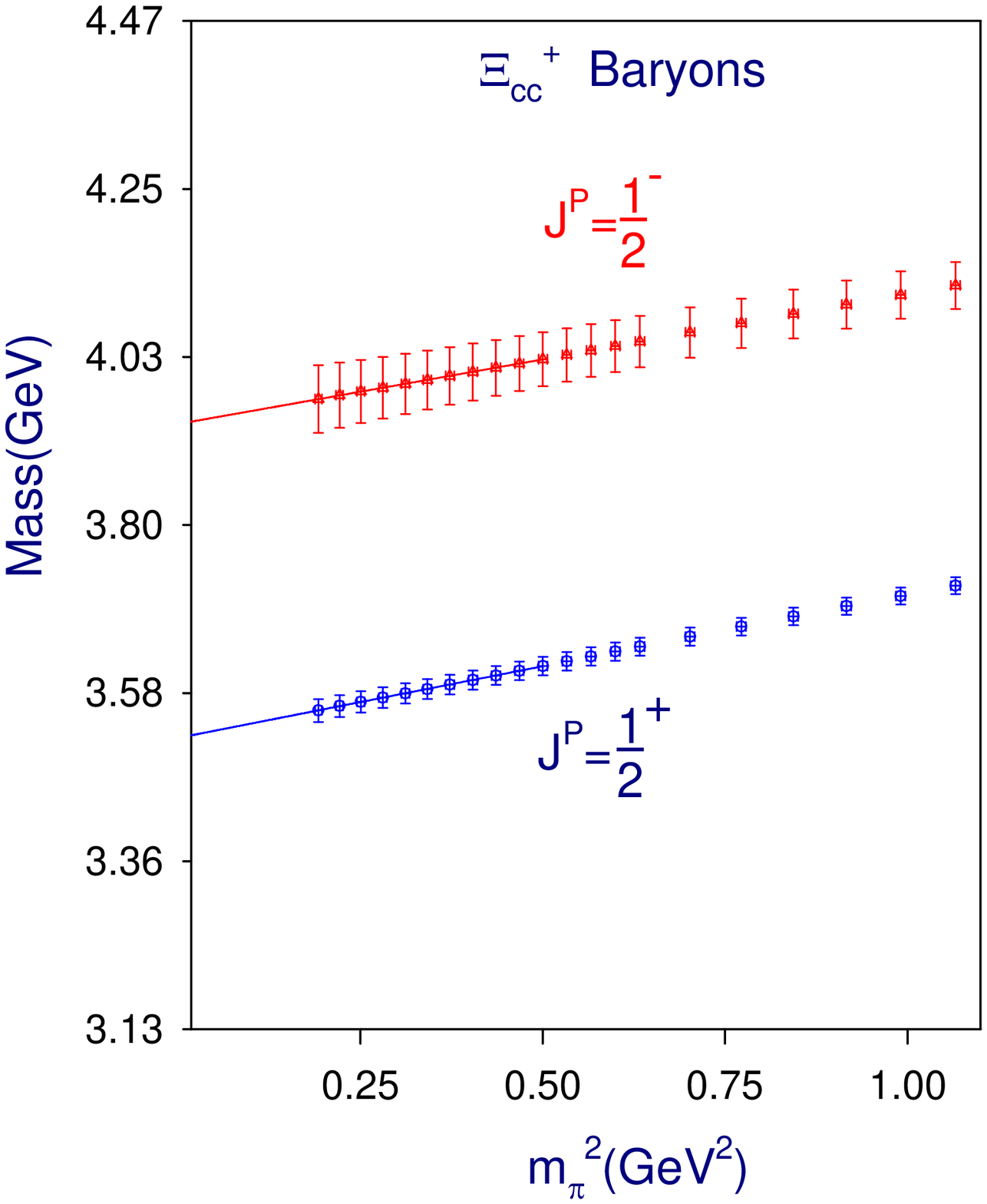}
\caption{The mass of the doubly charmed $ \Xi_{cc}^+ $ baryon 
versus the pion mass square,
for $ J^P = 1/2^\pm $ states respectively.
The solid lines are linear fits using the smallest 11 masses with
$ 0.03 \le m_d a \le 0.08 $. The even parity state
gives $ M_{\Xi_{cc}^+(1/2^+)} = 3522(16) $ MeV, in good agreement 
with $ \Xi_{cc}(3520) $ observed by SELEX \cite{Mattson:2002vu}.}
\label{fig:mXicc_20}
\end{minipage}
\end{figure}

One of the basic objectives of lattice QCD is to compute 
hadron masses nonperturbatively from the first principles. 
For hadrons only composed of strange and/or charm quarks 
(e.g., $ \Omega^0_c $ in Fig. \ref{fig:omega_c_20}),  
their masses (in quenched approximation) can be measured directly 
with presently accessible lattice sizes. 
However, for hadrons containing $ u, d $ light quarks,  
the performance of the present generation of computers is still
inadequate for computing their masses
at the physical scale ($ m_\pi \simeq 135 $ MeV), on a lattice
with enough number of sites in each direction such that the 
discretization errors and the finite volume effects both are
negligible comparing to the statistical ones.
Nevertheless, even with lattices of moderate sizes, lattice QCD
can determine the parameters of the hadron mass
formulas in the (quenched) chiral perturbation theory.
Then one can use these formulas to evaluate the hadron masses at the
physical scale. In this paper, we extrapolate the baryon mass 
linearly in $ m_\pi^2 $ (e.g., $ \Xi_{cc}^+ $ in Fig. \ref{fig:mXicc_20}). 

Here the quark fields are formulated  
in the framework of optimal domain-wall fermion  
\cite{Chiu:2002ir} such that the quark propagator and its fermion 
determinant can attain the maximal chiral symmetry for any finite $ N_s $.
For 100 gauge configurations generated with Wilson gauge action 
at $ \beta = 6.1 $ on the $ 20^3 \times 40 $ lattice, 
we compute (point-to-point) quark propagators  
for 30 quark masses in the range $ 0.03 \le m_q a \le 0.8 $      
\cite{Chiu:2003iw}.
Then we determine the inverse lattice spacing $ a^{-1} = 2.237(76) $ GeV 
from the pion correlation function, with the experimental input 
of pion decay constant $ f_\pi = 132 $ MeV. 
The strange quark bare mass $ m_s a = 0.08 $ 
and the charm quark bare mass $ m_c a = 0.8 $ are fixed such that 
the masses extracted from the vector meson time-correlation 
function are in close agreement with the experimental masses 
of $ \phi(1020) $ and $ J/\psi (3010) $ respectively. 
Then {\it the masses of any other hadrons
containing $ c, s, u $ and $ d $ quarks}\footnote{In this paper, we work in 
the isospin limit $ m_u = m_d $.} 
{\it are predictions of QCD from the first principles}. 

\section{Baryon interpolating operator and time-correlation fnnction}

First we define the notation for the ``diquark" operator
\BAN
[{\bf q^A} \Gamma {\bf q^B} ]_{xa} \equiv \epsilon_{abc} 
\Gamma_{\alpha\beta} 
( {\bf q^A}_{x\alpha b}{\bf q^B}_{x\beta c}
 -{\bf q^B}_{x\alpha b}{\bf q^A}_{x\beta c} )
\EAN
where ${\bf q^A}$ and ${\bf q^B}$ denote quark fields of  
flavors $ A $ and $ B $;  
$ \epsilon_{abc} $ is the completely antisymmetric tensor;
$ x $, $ \{ a, b, c \} $ and $ \{ \alpha, \beta, \gamma \} $
denote the lattice site, color, and Dirac indices respectively.
Here $ \Gamma_{\alpha\beta} = -\Gamma_{\beta\alpha} $ such 
that the diquark transforms like a spin singlet.
Further, we define
\BAN
({\bf q^A} \Gamma {\bf q^B} )_{xa} \equiv \epsilon_{abc} 
\Gamma_{\alpha\beta} {\bf q^A}_{x\alpha b}{\bf q^B}_{x\beta c}
\EAN
Then the interpolating operators for $N$ and $\Delta$ can be written as 
\BAN
N_{x\gamma} = [ {\bf u} (C\gamma_5) {\bf d} ]_{xa} {\bf u}_{x\gamma a}, 
\hspace{4mm}
\Delta_{x\gamma} = ({\bf d} C \gamma_\mu {\bf d})_{xa} {\bf d}_{x\gamma a}
\EAN
where $ C $ is the charge conjugation 
satisfying $ C \gamma_\mu C^{-1} = - \gamma_\mu^T $ and 
$ (C \gamma_5)^T = -C\gamma_5 $. 

Now suppressing the Dirac and site indices, 
the interpolating operators for baryons with $ J=1/2 $ can be 
written as 
\BAN
N \EQ [ \qu (C\gamma_5) \qd ] \qu,  \hspace{4mm} 
\Xi^0  = [ {\bf u} (C\gamma_5) {\bf s} ] {\bf s}, \hspace{4mm}   
\Sigma^+ = [ {\bf u} (C\gamma_5) {\bf s} ] {\bf u}  \\
\Lambda^0 \EQ [\qd (C\gamma_5) \qs] \qu + 
              [\qs (C\gamma_5) \qu] \qd - 2 [\qu (C\gamma_5) \qd] \qs, 
\EAN  
while the interpolating operators for baryons with 
$ J=3/2 $ can be written as
\BAN
\Delta^- \EQ (\qd C\gamma_\mu \qd) \qd, \hspace{4mm}   
\Omega^- = (\qs C\gamma_\mu \qs) \qs, \hspace{4mm}
\Sigma^+ = (\qu C\gamma_\mu \qs) \qu  + 
           (\qs C\gamma_\mu \qu) \qu  + (\qu C\gamma_\mu \qu) \qs,  \\   
\Xi^0 \EQ (\qu C\gamma_\mu \qs) \qs + 
          (\qs C\gamma_\mu \qs) \qu + (\qs C\gamma_\mu \qu) \qs   
\EAN  
Similarly, the operators for charmed baryons can be 
constructed, e.g.,  
$ \Omega_c^0 = [ {\bf c} (C\gamma_5) {\bf s} ] {\bf s}$, and
$ \Xi_{cc}^+ = [ {\bf c} (C\gamma_5) {\bf d} ] {\bf c} $.  

The time-correlation function of any baryon operator $ B $ is 
defined as 
$ 
C_{\alpha\beta}(t) = 
\sum_{\vec{x}} \langle B_{x\alpha} \bar B_{0\beta} \rangle
$
which can be expressed  
in terms of point-to-point quark propagators from 
$ y=(\vec{0},0) $ to $ x=(\vec{x},t) $.  
Its average over gauge configurations is fitted by the usual formula 
\BAN
\langle C(t) \rangle = 
\frac{1+\gamma_4}{2} ( Z_{+} e^{-m_{+} a t}
                              -Z_{-} e^{-m_{-}a (T-t)}) 
   +\frac{1-\gamma_4}{2} ( Z_{+} e^{-m_{+}a (T-t)}
                              -Z_{-} e^{-m_{-}a t})
\EAN   
where $ m_\pm $ are the masses of even and odd parity states. 
Thus, one can use parity projector $ ( 1 \pm \gamma_4 )/2 $
to project out two amplitudes,
\BAN 
A_{+}(t) \equiv Z_{+} e^{-m_{+}a t} -Z_{-} e^{-m_{-}a (T-t)},  \hspace{2mm}
A_{-}(t) \equiv Z_{+} e^{-m_{+}a (T-t)} -Z_{-} e^{-m_{-}a t}. 
\EAN
Now the problem is how to extract $ m_{\pm} $ from $ A_{\pm} $ respectively. 
Obviously, for sufficiently large $ T $, there exists 
a range of $ t $ such that, in $ A_{\pm} $, the contributions due to 
the opposite parity state are negligible. 
Then $ m_{\pm} $ can be extracted by a single exponential fit to $ A_{\pm} $,  
for the range of $ t $ in which the effective mass 
$ m_{eff}(t) = \ln(A_{\pm}(t)/A_{\pm}(t+1)) $ attains a plateau.    
On the other hand, if $ T $ is not so large, then it may turn out 
that the heavier mass, say $ m_- $ (assuming $ m_- > m_+ $), could not be 
easily extracted from $ A_- $ due to the non-negligible contributions of 
the (lowest lying) even parity state. 
For our lattice with $ T = 40 $, it is sufficiently large to extract 
the mass of the lowest lying state for the entire range of 
$ 0.03 \le m_q a \le 0.8 $.    
However, the excited state seems to suffer from the  
(backward propagating) contribution of  
the lowest-lying state, especially for $ m_q a \le 0.05 $,   
which yields relatively larger error in extracting the mass of the 
negative parity state.

\section{Summary and Concluding Remarks}

Our results of baryon mass spectra are summarized in
Tables \ref{tab:light_baryon_mass} and \ref{tab:charm_baryon_mass},
along with the experimental mass spectra listed by the
Particle Data Group \cite{Eidelman:2004wy}.
The empty entries in the last column are baryons which  
have not been discovered in experiments.
Thus the baryon masses obtained in this paper
can serve as theoretical predictions of lattice QCD, 
in particular, the singly-charmed, doubly-charmed, 
and triply-charmed baryons. Details of our results
including meson mass spectra will be presented elsewhere \cite{Chiu:2005hm}.
Our results of baryon masses are in good agreement with experimental
values, except for some of the negative parity baryons,  
in particular, $ \Lambda(1405) $ and $ \Lambda_c(2593) $. 
This certainly warrants further studies   
on these excited baryons 
(e.g., with higher statistics, larger lattices, smaller $ u(d) $ quark masses, 
and incorporation of dynamical fermions). 


\begin{table}[htb]
\begin{minipage}[t]{78mm}
\caption{
Baryon mass spectra obtained in this paper
[with $ m_s a = 0.08 $, $ m_c a = 0.80 $ and $ a^{-1} = 2237(76) $ MeV].
The last column is from the  
listings of Particle Data Group \cite{Eidelman:2004wy}, 
where $ J^P $ has not been measured for all entries.
}
\begin{tabular}{c|cccc}
Baryon         & $ J^P $ & Mass & Expt \\
\hline
\hline
$ N $          & $ 1/2^+ $   & 958(36)   &  939  \\
$ N $          & $ 1/2^- $   & 1603(150) &  1535  \\
$ \Delta $     & $ 3/2^+ $   & 1243(74)  &  1232  \\
$ \Delta $     & $ 3/2^- $   & 1764(105) &  1700  \\
$ \Lambda $    & $ 1/2^+ $   & 1119(38)  &  1116  \\
$ \Lambda $    & $ 1/2^- $   & 1854(47)  &  1405  \\
$ \Sigma $     & $ 1/2^+ $   & 1211(43)  &  1189  \\
$ \Sigma $     & $ 1/2^- $   & 1872(118) &  1750  \\
$ \Xi $        & $ 1/2^+ $   & 1326(32)  &  1315  \\
$ \Xi $        & $ 1/2^- $   & 1935(110) &  1950  \\
$ \Sigma $     & $ 3/2^+ $   & 1376(61)  &  1385  \\
$ \Sigma $     & $ 3/2^- $   & 1931(93)  &  1940  \\
$ \Xi $        & $ 3/2^+ $   & 1539(42)  &  1530  \\
$ \Xi $        & $ 3/2^- $   & 2039(95)  &  2030  \\
$ \Omega $     & $ 3/2^+ $   & 1668(29)  &  1672  \\
$ \Omega $     & $ 3/2^- $   & 2229(50)  &  2250  \\
\hline
$ \Omega_{ccc} $   & $ 3/2^+ $   & 4681(28) &     \\
$ \Omega_{ccc} $   & $ 3/2^- $   & 5066(48) &     \\
\hline
\end{tabular}
\label{tab:light_baryon_mass}
\end{minipage}
\hspace{\fill}
\begin{minipage}[t]{77mm}
\caption{Continuation of Table \ref{tab:light_baryon_mass}.
}
\begin{tabular}{c|ccc}
   Baryon          & $ J^P $ & Mass & Expt. \\
\hline
\hline
$ \Lambda_{c} $    & $ 1/2^+ $   & 2275(27) &  2285  \\
$ \Lambda_{c} $    & $ 1/2^- $   & 3027(63) &  2593  \\
$ \Sigma_{c} $     & $ 1/2^+ $   & 2458(24) &  2455  \\
$ \Sigma_{c} $     & $ 1/2^- $   & 2882(40) &        \\
$ \Xi_{c} $        & $ 1/2^+ $   & 2478(27) &  2466  \\
$ \Xi_{c} $        & $ 1/2^- $   & 2793(68) &  2790  \\
$ \Omega_{c} $     & $ 1/2^+ $   & 2677(30) &  2697  \\
$ \Omega_{c} $     & $ 1/2^- $   & 3100(130)&        \\
$ \Sigma_{c} $     & $ 3/2^+ $   & 2522(30) &  2520  \\
$ \Sigma_{c} $     & $ 3/2^- $   & 3007(47) &        \\
$ \Xi_{c} $        & $ 3/2^+ $   & 2646(36) &  2645  \\
$ \Xi_{c} $        & $ 3/2^- $   & 2954(94) &  2815  \\
$ \Omega_{c} $     & $ 3/2^+ $   & 2756(32) &        \\
$ \Omega_{c} $     & $ 3/2^- $   & 3224(62) &        \\
\hline
$ \Xi_{cc} $       & $ 1/2^+ $   & 3522(16) &  3520  \\
$ \Xi_{cc} $       & $ 1/2^- $   & 3940(47) &        \\
$ \Omega_{cc} $    & $ 1/2^+ $   & 3637(23) &        \\
$ \Omega_{cc} $    & $ 1/2^- $   & 4053(57) &        \\
$ \Xi_{cc} $       & $ 3/2^+ $   & 3655(20) &        \\
$ \Xi_{cc} $       & $ 3/2^- $   & 4043(22) &        \\
$ \Omega_{cc} $    & $ 3/2^+ $   & 3762(17) &        \\
$ \Omega_{cc} $    & $ 3/2^- $   & 4147(31) &        \\
\hline
\end{tabular}
\label{tab:charm_baryon_mass}
\end{minipage}
\end{table}

\end{document}